\documentclass[prd,aps,a4paper,nofootinbib]{revtex4}  

\newif\ifusesec
\usesectrue  
   
\usepackage{graphicx} 
\usepackage{mathrsfs}
\usepackage{amsmath,amsfonts,amssymb}
\usepackage{multirow}

\newcommand{\beq}{\begin{equation}}
\newcommand{\eeq}{\end{equation}}
\newcommand{\bea}{\begin{eqnarray}}
\newcommand{\eea}{\end{eqnarray}}

\begin{document}

\title{Scattering of a point mass by a Schwarzschild black hole: radiated energy and angular momentum}

\author{Andrea Geralico}
\affiliation{
Istituto per le Applicazioni del Calcolo ``M. Picone,'' CNR, I-00185 Rome, Italy
}

\date{\today}

\begin{abstract}
The radiated energy and angular momentum from a point mass on a hyperbolic-like orbit about a Schwarzschild black hole are computed for the first time in the framework of the first-order self-force theory.
The analytical expressions for the fluxes are obtained through the standard method of Mano, Suzuki and Takasugi in the form of combined post-Minkowskian (PM) and post-Newtonian (PN) expansions.
The reached PM accuracy for the energy and angular momentum losses is $O(G^5)$ and $O(G^4)$, respectively, and 7PN for both.
The radiative losses (energy, angular momentum and linear momentum) are currently known in PN-expanded form up to the (fractional) 3PN order [D. Bini et al., Phys. Rev. D \textbf{107}, 024012 (2023)]. 
Exact PM results valid for arbitrary values of the velocity are limited to $O(G^4)$ for the energy and $O(G^3)$ for the angular momentum.
An exact expression for the radiated energy at $O(G^5)$ has been recently obtained in [M. Driesse et al., Nature \textbf{641}, 603-607 (2025)] in the first-order self force limit, whereas the $O(G^4)$ radiated angular momentum has been partially determined 
in [C. Heissenberg, Phys. Rev. D \textbf{111}, 126012 (2025)].
The results of this work are in agreement with the state of the art of both energy and angular momentum losses, also completing the knowledge of the 4PM radiated angular momentum up to the reached PN accuracy.
The expressions for radiative losses are then used to get the 5PM radiation-reacted scattering angle, which should also serve as a cross-check of ongoing calculations by other methods.
\end{abstract}


\maketitle

\section{Introduction}

Analytic self-force (SF) calculations so far have been mainly focussed on bound-orbit configurations, which are relevant for the modelling of extreme mass-ratio inspiral systems.
All techniques developed in this context are based on a frequency domain approach to the solution of the perturbation equations and the metric reconstruction procedure, which strongly depends on the periodicity of the orbit leading to a discrete spectrum of frequencies \cite{Barack:2018yvs}.

The primary goal of gravitational self-force theory is the computation of gauge-invariant quantities, which allows for cross-checking of results obtained through different analytical approximations schemes, including post-Minkowskian (PM) \cite{Damour:2016gwp,Damour:2017zjx,Bini:2017xzy,Vines:2017hyw,Bini:2018ywr,Bjerrum-Bohr:2019kec,Kalin:2019inp,Kalin:2020mvi,Damour:2019lcq,Mogull:2020sak}, post-Newtonian (PN) \cite{Blanchet:2013haa,Bini:2017wfr}, effective-one-body (EOB) \cite{Buonanno:1998gg,Damour:2000we}, effective-field-theory (EFT) \cite{Levi:2018nxp,Foffa:2013qca,Foffa:2021pkg,Kalin:2020fhe,Dlapa:2021npj,Dlapa:2021vgp}, and Tutti Frutti (TF) \cite{Bini:2019nra,Bini:2020wpo,Bini:2020nsb,Bini:2020hmy,Bini:2020rzn}. 
The information provided by all these approaches leads to a continuously improving knowledge of the dynamics of binary systems and of the associated gravitational wave emission.

In recent years the attention of the gravity community has been drawn by scattering processes due to development of new powerful methods such as generalized unitarity and double copy \cite{Cheung:2018wkq,Bern:2019nnu,Bern:2019crd,Bjerrum-Bohr:2018xdl,Kosower:2018adc,Herrmann:2021lqe}, allowing for translating between quantum scattering amplitudes and classical gravitational dynamics in a PM framework.
The state of the art in the case of spinless bodies is 4PM for both the conservative scattering and the dissipative case including radiation-reaction effects \cite{Bern:2021dqo,Bern:2021yeh,Dlapa:2022lmu,Dlapa:2024cje}.
Recent works have also initiated the study of the 5PM scattering dynamics in the first-order self-force (1SF) limit \cite{Driesse:2024xad,Driesse:2024feo,Dlapa:2025biy}.

When considering an unbound motion in the SF framework the main issue to be addressed is how to handle a continuous spectrum of frequencies (see Refs. \cite{Hopper:2017qus,Hopper:2017iyq} as the first studies in this direction).
The analytical setup needed to perform a SF computation in the hyperbolic-like case has been recently discussed in Ref. \cite{Barack:2022pde}, even if in the simplest case of a scalar charge scattered by a Schwarzschild black hole and implemented only numerically to obtain the estimates of both the conservative and dissipative SF corrections to the scattering angle.
A consistency check for these results has been provided very recently by Ref. \cite{Barack:2023oqp}, where the scattering-amplitude methods have been applied to this problem to get (exact) PM analytical expressions for the scattering angle (with 4PM accuracy for both conservative and dissipative parts) and the (3PM) radiated energy and (2PM) angular momentum.
At the 4PM order the conservative scattering angle has two undetermined coefficients related to the way to model the contributions from tidal effects, which have been estimated upon comparison with the numerical SF results of Ref. \cite{Barack:2022pde}. These coefficients have been then fixed in Ref. \cite{Bini:2024icd}, which provides the first analytical computation for unbound motion within the SF approach even if in the simplest scalar case. 
However the strategy presented there is general enough to be applied to the more interesting situation of pure gravitational scattering, in which case however additional delicate issues should be addressed.

The aim of this work is to compute the radiated energy and angular momentum emitted by a point mass scattered off a Schwarzschild black hole in the first-order self-force formalism. 
When considering the scattering of two bodies with masses $m_1$ and $m_2$ it is convenient to parametrize the PM expansions of the radiative losses by the coefficients of their power series expansion in the inverse dimensionless angular momentum $\frac1j= \frac{G m_1 m_2}{cJ}$ as follows
\bea
\label{various_expansions}
\frac{E^{\rm rad}}{M} &=&  \nu^2   \sum_{n=3}^\infty \frac{ E_{n}}{j^n}\,,\nonumber\\
\frac{ J^{\rm rad}}{J_{\rm c.m.}} &=& \nu  \sum_{n=2}^\infty \frac{{ J}_{n}}{j^n} \,, 
\eea
where $\nu=m_1m_2/M^2$ is the symmetric mass ratio, $M=m_1+m_2$ the total mass, $J_{\rm c.m.} = b \mu p_\infty/h=GM^2\nu j$ the center-of-mass angular momentum, and $h=E/Mc^2\equiv\sqrt{1+2\nu(\gamma-1)}$ the dimensionless incoming center-of-mass energy  (with $p_\infty=\sqrt{\gamma^2-1}$).
Recalling that $j$ is related to the impact parameter $b$ by $\frac1j= \frac{G M h}{ b p_\infty}$, any expansion in powers of $1/j$ corresponds to an expansion in powers of $GM/b$, i.e., to a PM expansion.
The coefficients $E_n$ and $J_n$ are functions of $\gamma$ and $\nu$, and are mostly known in a PN-expanded form (i.e., for small values of $p_\infty$). Their expressions up to $n=7$ accurate to the 3PN order can be found, e.g., in Refs. \cite{Bini:2021gat,Bini:2022enm}. 
The exact values of the 3PM and 4PM coefficients $E_3$ and $E_4$ have been computed in Refs. \cite{Bern:2021dqo,Herrmann:2021lqe,Herrmann:2021tct} and \cite{Dlapa:2022lmu}, respectively.
The 2PM and 3PM coefficients $J_2$ and $J_3$ are also known exactly (see Refs. \cite{Damour:2020tta} and \cite{Manohar:2022dea}, respectively).

Further progress in the determination of these coefficients at higher PM orders has been made very recently in Ref. \cite{Driesse:2024feo}, where the energy coefficient $E_5$ has been computed in the 1SF limit (i.e., for $\nu\to0$) by using the world line quantum field theory formalism.
The angular momentum coefficient $J_4$, instead, has been partially computed in Ref. \cite{Heissenberg:2025ocy} through amplitude-based methods. In fact, the expression obtained there is exact as a function of $\gamma$, but its small-velocity expansion gives only the even powers of $p_\infty$ (corresponding to PN fractional orders). 
The contribution of the present paper is twofold: it provides the first check of the results of Ref. \cite{Driesse:2024feo} up to the 7PN order, and most importantly confirms and complete those of Ref. \cite{Heissenberg:2025ocy} with the same PN accuracy.
Furthermore, the knowledge of the radiative losses allows for determining the 5PM-1SF contribution to the radiation-reacted scattering angle, whose general PM expansion reads 
\beq
\label{chiraddef}
\delta \chi^{\rm rad} = \sum_{n=3}^\infty \frac{ 2\chi^{\rm rad}_{n}}{j^n}\,.
\eeq
The addition of the conservative part will give the total scattering angle, which has also been computed in Ref. \cite{Driesse:2024feo}, representing then a second independent check of their results.
However, the conservative scattering angle requires a separate calculation, which is left for future work.

\section{Self-force corrections to the scattering angle}

Let us consider a point particle with mass $m_1$ moving along a hyperbolic-like orbit on the equatorial plane ($\theta=\frac{\pi}{2}$) of a Schwarzschild black hole with mass $m_2$ (with $m_1\ll m_2$) under the effect of its self-force.
The equations of motion then read
\beq
\label{geo_eqs}
\frac{du_\alpha}{d\tau}-\frac12 g_{\mu\nu,\alpha}u^\mu u^\nu=F_\alpha\,,
\eeq
where $g_{\mu\nu}=\bar g_{\mu\nu}+h_{\mu\nu}$ and $u^\alpha=\bar u^\alpha+\delta u^\alpha$ are the perturbed metric and the perturbed four-velocity, respectively, and
\beq
F^\mu=-\frac12(\bar g^{\mu\nu}+\bar u^\mu\bar u^\nu)\bar u^\lambda\bar u^\rho(2h_{\nu\lambda;\rho}-h_{\lambda\rho;\nu})\,,
\eeq
the self-force per unit particle's mass, a bar denoting the corresponding background quantities.
All quantities $h_{\mu\nu}$, $\delta u^\alpha$ and $F^\mu$ are first order in the mass ratio $q=\frac{m_1}{m_2}$ (or equivalently in $\nu=q+O(q^2)$), so that for instance the $4$-velocity 1-form field is $u_\alpha =\bar u_\alpha +h_{0\alpha}+\bar g_{\alpha\beta}\delta u^\beta$ to first order in $q$.

It is convenient to introduce the first order quantities $\delta E=-\bar g_{t\alpha} \delta u^\alpha$ and $\delta L = \bar g_{\phi\alpha} \delta u^\alpha$ such that the four velocity components $u^\alpha$ can be written exactly in the same form as those of the background with the replacement $\bar E\to \bar E+\delta E$ and $\bar L\to \bar L+\delta L$, implying that $\delta u_t =-\delta E$ and $\delta u_\phi = \delta L$.
The correction $\delta u^r$ to the radial component of the four velocity directly follows from the normalization condition of $u$ ($u\cdot u=-1$) with respect to the perturbed metric, which reads
\beq
\bar g_{rr} \bar u^r \delta u^r = \bar u^t \delta E- \bar u^\phi \delta L-\frac12 h_{00}\,,
\eeq
where $h_{00}=h_{\alpha\beta}\bar u^\alpha \bar u^\beta$.

Equations  \eqref{geo_eqs} thus determine the evolution of $\delta u_t$ and $\delta u_\phi$, or equivalently of the perturbations in energy and angular momentum by
\bea
\label{eqdeltaEL_BS}
\frac{d}{d\tau}\hat \delta E = -F_t \,,\qquad 
\frac{d}{d\tau}\hat \delta L = F_\phi\,,
\eea
where (see Ref. \cite{Barack:2011ed}, where hatted quantities satisfying the normalization condition of $u$ with respect to the background metric has been conveniently introduced, so that $\bar g_{rr} \bar u^r \hat\delta u^r = \bar u^t \hat\delta E- \bar u^\phi \hat\delta L$) 
\bea
\label{relwithBS}
\hat\delta E&=&\delta E-\frac12\bar E h_{00}\,, \nonumber\\
\hat\delta L&=&\delta L-\frac12\bar L h_{00}\,,
\eea
which can be formally integrated as
\beq
\hat\delta E(\tau)=-\int_{-\infty}^\tau F_t d\tau\,,\qquad 
\hat\delta L(\tau)=\int_{-\infty}^\tau F_\phi d\tau\,,
\eeq
having assumed the initial conditions
\beq
\hat\delta E(\tau=-\infty)=0=\hat\delta L(\tau=-\infty)\,.
\eeq
The self-force can be split into conservative and dissipative pieces $F_\alpha=F_\alpha^{\rm cons}+F_\alpha^{\rm diss}$ in terms of the retarded and advanced metric perturbations \cite{Barack:2010tm}, according to 
\bea
\label{Fconsdissdef}
F_\alpha^{\rm cons}=\frac12[F_\alpha(\tau)\pm F_\alpha(-\tau)]
\,,\nonumber\\
F_\alpha^{\rm diss}=\frac12[F_\alpha(\tau)\mp F_\alpha(-\tau)]\,,
\eea
with the upper sign for $\alpha=r$ and the lower sign for $\alpha=t,\phi$.
The dissipative part is responsible for the removal of energy and angular momentum from the system.

The scattering angle is defined as 
\beq
\chi=\int_{-\infty}^{\infty}d\tau \frac{d\phi}{d\tau}-\pi\,,
\eeq
and can be conveniently computed by using the radial coordinate as a parameter along the orbit \cite{Bini:2024icd}, i.e.,
\beq
\chi=\sum_\pm\int_{r_{\rm min}}^\infty dr \left(\frac{d\phi}{dr}\right)^\pm-\pi\,.
\eeq
where the label $\pm$ denotes the value of the various quantities corresponding to the incoming ($-$, $\bar u^r<0$) and outgoing ($+$, $\bar u^r>0$) branches, with $r$ decreasing from infinity ($\tau \to-\infty$) up to a minimum value $r_{\rm min}=\bar r_{\rm min}+\delta r_{\rm min}$ ($\tau=0$), then increasing again to infinity ($\tau \to\infty$).

To first order in $q$ the integrand reads
\bea
\frac{d\phi}{dr}&=& \frac{\bar u^\phi+\delta u^\phi}{\bar u^r+\delta u^r}\nonumber\\
&=& \frac{\bar u^\phi}{\bar u^r}\left(1+\frac{\delta u^\phi}{\bar u^\phi}-\frac{\delta u^r} {\bar u^r}\right)\nonumber\\
&\equiv& \frac{d\bar \phi}{dr}+\frac{d\,\delta\phi}{dr}\,,  
\eea
where
\beq
\label{eqdeltaphi}
\frac{d\,\delta\phi}{dr} =a_E(r)\hat\delta E(r)+a_L(r)\hat\delta L(r)\,,
\eeq
with
\beq
a_E(r)=-\frac{\bar L \bar E}{r^2(\bar u^r)^3 }
\,,\qquad
a_L(r)=\frac{\bar E^2-f(r)}{r^2(\bar u^r)^3 }\,,
\eeq
and $f(r)=1-2m_2/r$.

Integration of the geodesic term $\frac{d\bar \phi}{dr}$ gives a first-order correction due to the fact that the minimum approach also changes. The result of the integration is then formally equal to the geodesic value, but with $\bar E\to \bar E+\hat\delta E_0$ and $\bar L\to \bar L+\hat\delta L_0$, with $\hat\delta E_0=\hat\delta E(\tau=0)$ and $\hat\delta L_0=\hat\delta L(\tau=0)$, so that 
\beq
\chi=\bar\chi+\delta\bar\chi+\hat\delta\phi\,,
\eeq
where $\bar\chi=\bar\chi(\bar E,\bar L)$ is the geodesic value of the scattering angle (which is known exactly in terms of Elliptic integrals), with first-order corrections
\beq
\delta\bar\chi=\frac{\partial\bar\chi}{\partial \bar E}\hat\delta E_0+\frac{\partial\bar\chi}{\partial \bar L}\hat\delta L_0\,,
\eeq
and 
\beq
\hat\delta\phi=\sum_\pm\int_{\bar r_{\rm min}}^\infty dr \left[a_E^\pm(r)\hat\delta E^\pm(r)+a_L^\pm(r)\hat\delta L^\pm(r)\right]\,,
\eeq
with
\beq
a_E^-(r)=-a_E^+(r)\,,\qquad
a_L^-(r)=-a_L^+(r)\,.
\eeq

The conservative and dissipative components \eqref{Fconsdissdef} of the self-force satisfy the following properties
\bea
F_\alpha^{\rm cons\,+}(r)&=&\frac12[F_\alpha^+(r)-F_\alpha^-(r)]
=-F_\alpha^{\rm cons\,-}(r)
\,,\nonumber\\
F_\alpha^{\rm diss\,+}(r)&=&\frac12[F_\alpha^+(r)+F_\alpha^-(r)]
=F_\alpha^{\rm diss\,-}(r)\,,
\eea
for $\alpha=t,\phi$, so that 
\beq
\hat\delta\phi^{\rm cons}=-2\int_{\bar r_{\rm min}}^{\infty} dr \left[
a_E^+(r)\int_{\bar r_{\rm min}}^r\frac{dr}{\bar u^r}F_t^{\rm cons\,+}
-a_L^+(r)\int_{\bar r_{\rm min}}^r\frac{dr}{\bar u^r}F_\phi^{\rm cons\,+}
\right]
\,,
\eeq
whereas $\hat\delta\phi^{\rm diss}=0$.
In the dissipative case the SF correction to the scattering angle is then simply given by (note the close analogy with the linear response formula \cite{Bini:2012ji,Damour:2020tta})
\beq
\delta\chi^{\rm diss}=\delta\bar\chi^{\rm diss}=-\frac12\left(\frac{\partial\bar\chi}{\partial \bar E}E_{\rm rad}+\frac{\partial\bar\chi}{\partial \bar L}L_{\rm rad}\right)\,,
\eeq
where $E_{\rm rad}=-2\hat\delta E_0^{\rm diss}$ and $L_{\rm rad}=-2\hat\delta L_0^{\rm diss}$ are the total radiated energy and angular momentum, respectively.
It is worth noting that $E_{\rm rad}$ and $L_{\rm rad}$ here come from the local dissipative force, so that they also include the energy and angular momentum losses into the black hole horizon, i.e., $E_{\rm rad}=E_{\rm rad}^H+E_{\rm rad}^\infty$ and $L_{\rm rad}=L_{\rm rad}^H+L_{\rm rad}^\infty$.

Therefore, in order to calculate the radiation-reacted scattering angle 
\beq
\label{chirad}
\chi^{\rm rad}=-\frac12\left(\frac{\partial\bar\chi}{\partial \bar E}E_{\rm rad}^\infty+\frac{\partial\bar\chi}{\partial \bar L}L_{\rm rad}^\infty\right)\,,
\eeq
one needs the radiated energy and angular momentum at infinity, whose computation by directly integrating the fluxes at infinity is discussed in the next section.

\section{Gravitational radiation reaction for unbound orbits}

According to the Teukolsky formalism the information on the radiation emitted by the system is encoded in the Weyl scalar $\psi_4$, which is asymptotically related to the two independent polarizations $h_+$ and $h_\times$ of the gravitational waves by
\beq
\label{hdef}
\psi_4(r\to\infty)\sim-\frac12(\ddot h_+-i\ddot h_\times)\equiv-\frac12\ddot h\,,
\eeq
where a dot denotes time derivative.
The particle is assumed to move along an hyperboliclike geodesic orbit on the equatorial plane of a Schwarzschild spacetime with parametric equations $x^\mu =x_p^{\mu}(\tau)$.
The associated energy momentum tensor is proportional to a three-dimensional Dirac delta function with support on the particle position.

The Weyl scalar $\psi_4$ satisfies the Teukolsky equation with spin-weight $s=-2$. 
Separation of variables and Fourier-transforming gives
\beq
\label{sep}
\psi_4= \frac1{r^4}\int\frac{d\omega}{2\pi}e^{-i\omega t}\sum_{lm}\,\,R_{lm\omega}(r)\,\, {}_{-2}Y_{lm}(\theta,\phi)\,,
\eeq
where ${}_{s}Y_{lm}(\theta,\phi)$ are spin-weighted spherical harmonics.
The radial function $R_{lm\omega}(r)$ satisfies the inhomogeneous equation 
\beq
\label{radeq}
\left\{\Delta^{2}\frac{d}{dr} \left(\Delta^{-1} \frac{d}{dr} \right) +\left[\frac{K^2+4i (r-m_2)K}{\Delta}-8i\omega r -\lambda\right]\right\}R_{lm\omega }(r)=-8\pi T_{lm\omega}(r)\,,
\eeq
where $\Delta=r(r-2m_2)$, $K=r^2\omega$ and $\lambda=(l-1)(l+2)$, and $T_{lm\omega}(r)$ is the harmonic decomposition of the source term.
Eq. \eqref{radeq} can be solved by using the Green's function method in terms of the ingoing and upgoing homogeneous solutions $R^{\rm in}_{lm\omega}(r)$ and $R^{\rm up}_{lm\omega}(r)$, having the correct behavior at the horizon and at infinity, respectively.
The asymptotic solution representing purely outgoing waves is given by 
\beq
R_{lm\omega }(r\to\infty)\sim Z^\infty_{lm\omega} r^3e^{i\omega r_*}\,,
\eeq
where $r_*$ is the tortoise coordinate, and $Z^\infty_{lm\omega}$ is the amplitude
\beq
\label{Zinf}
Z^\infty_{lm\omega}=\frac{C^{\rm trans}_{lm\omega}}{W_{lm\omega}}\int_{2m_2}^\infty dr\frac{R^{\rm in}_{lm\omega}(r)T_{lm\omega}(r)}{\Delta^2}\,,
\eeq
with $W_{lm\omega}$ the (constant) Wronskian, and $C^{\rm trans}_{lm\omega}$ the (constant) transmission coefficient.
This is a well established framework.
Full details can be found, e.g., in Ref. \cite{Sasaki:2003xr}, to which I also refer for notation and conventions. 

The asymptotic form of $\psi_4$ then turns out to be
\beq
\label{sep2}
\psi_4= \frac1{r}\sum_{lm}\int\frac{d\omega}{2\pi}Z^\infty_{lm\omega}e^{-i\omega u}\,\,{}_{-2}Y_{lm}(\theta,\phi)\,,
\eeq
where $u=t-r_*$ is the retarded time, and the final expression for the amplitude \eqref{Zinf} has the form
\beq
\label{Zinf2}
Z^\infty_{lm\omega}=\frac{C^{\rm trans}_{lm\omega}}{W_{lm\omega}}\int dt'e^{i\omega (t'-m\phi_p(t'))}F_{lm\omega}(r_p(t'))\,.
\eeq
The function $F_{lm\omega}(r_p(t))$ is the result of the action of a differential operator on the ingoing homogeneous solution $R^{\rm in}_{lm\omega}(r)$ upon evaluation on the particle position $r=r_p(t)$. 

The 1SF energy and angular momentum fluxes are given by
\bea
\label{fluxes}
\frac{dE_{\rm rad}^\infty}{du} &=&  \frac{r^2}{16\pi}\int d\Omega|\dot h|^2\,,\nonumber\\
\frac{dJ_{\rm rad}^\infty}{du} &=&  -\frac{r^2}{16\pi}{\rm Re}\left[\int d\Omega (\partial_\phi h)\bar{\dot h}\right]\,, 
\eea
where $d\Omega=\sin\theta d\theta d\phi$ is the standard solid angle element, and $\dot h$ and $h$ are obtained by integrating $\psi_4$, Eq. \eqref{sep2}, with respect to $u$ once and twice, respectively, according to Eq. \eqref{hdef}, so that
\beq
\label{doth}
\dot h= \frac2{r}\sum_{lm}\int\frac{d\omega}{2\pi}\frac{Z^\infty_{lm\omega}}{i\omega}e^{-i\omega u}\,\,{}_{-2}Y_{lm}(\theta,\phi)\,,
\eeq
and
\beq
\label{h}
h= \frac2{r}\sum_{lm}\int\frac{d\omega}{2\pi}\frac{Z^\infty_{lm\omega}}{\omega^2}e^{-i\omega u}\,\,{}_{-2}Y_{lm}(\theta,\phi)\,.
\eeq

The amplitude \eqref{Zinf2} is computed by using the MST solutions, $R^{\rm in(MST)}_{lm\omega}(r)$ and $R^{\rm up(MST)}_{lm\omega}(r)$, satisfying the retarded boundary conditions of ingoing radiation at the horizon and upgoing at infinity \cite{Mano:1996mf,Mano:1996vt}.
In the case of bound motion integrating over time in Eq. \eqref{Zinf2} simply gives a sum of Dirac-delta functions of the frequency, since the spectrum contains only a discrete set of distinct frequencies, i.e., the harmonics of the fundamental frequencies of the motion, implying that further integration over frequencies, Eqs. \eqref{doth}--\eqref{h}, is straightforward.
In the case of unbound motion characterized by a continuous spectrum instead it is convenient to integrate over the frequencies first, as discussed in Ref. \cite{Bini:2024icd}.
This step requires some care. 
In fact, the MST solutions are not only polynomial functions of the frequency just as the PN solutions, but contain also logarithmic terms of the type $\omega^n\ln^k\omega$.
In the former case the integration is done straightforwardly in terms of the Dirac-delta function and its derivatives according to the rule $\omega^n\to i^{-n}\delta^{(n)}(t'-t)$.
In the latter case instead the relevant integrals are of the type \cite{Bini:2024icd,DiRusso:2025lip}
\beq
I_k=\int \frac{d\omega }{2\pi}e^{i\omega(t'-t)}\ln^k\omega
=\frac{A_k}{t'-t}H(t-t')+B_k\delta(t'-t)\,,
\eeq
where $H(x)$ denotes the Heaviside step function, and the coefficients $A_k$ are polynomial functions of $\ln^{k-1}(t-t')$.
Further integration over $t'$ of the the first term leads to integrals of the type 
\beq
\label{nlint}
\int_{-\infty}^tdt'\,\frac{F(t')\ln^{k-1}(t-t')}{t'-t}\,,
\eeq
which diverge for $t'=t$, and are evaluated by taking their Hadamard finite part.
Those terms with $k>1$ are responsible for the appearance of multiple polylogarithms starting from the 6PM level for the energy and 5PM for the angular momentum.

\section{Results}

I list below the 1SF contributions to the energy and angular momentum losses computed in this work, i.e., 
\bea
E_5&=&E_5^{\rm 1SF}+O(\nu)\,,\nonumber\\
J_4&=&J_4^{\rm 1SF}+O(\nu)\,,
\eea
which are both accurate at the 7PN order. 

The 5PM-1SF radiated energy turns out to be 
\bea
E_5^{\rm 1SF}&=&\pi\left[
\frac{122}{5} p_\infty^2
+\frac{13831}{280} p_\infty^4
+\frac{297\pi^2}{20} p_\infty^5
-\frac{64579}{5040} p_\infty^6
+\left(\frac{9216}{35}-\frac{24993 \pi ^2}{1120}\right) p_\infty^7\right.\nonumber\\
&&
+\left(-\frac{10593 \ln(p_\infty/2)}{350}+\frac{99 \pi^2}{10}+\frac{29573617463}{310464000}\right) p_\infty^8
+\left(\frac{1577 \pi^2}{1120}-\frac{384}{7}\right) p_\infty^9\nonumber\\
&&
+\left(\frac{398799 \ln(p_\infty/2)}{19600}-\frac{3 \pi^2}{112}-\frac{1637965903901}{18834816000}\right) p_\infty^{10}
+\left(\frac{68478016}{202125}-\frac{883321 \pi^2}{22528}\right) p_\infty^{11}\nonumber\\
&&
+\left(-\frac{39476531
   \ln(p_\infty/2)}{705600}+\frac{40441 \pi^2}{3360}+\frac{120280025794939}{508540032000}\right) p_\infty^{12}
+\left(\frac{761964843 \pi^2}{41000960}-\frac{8048589184}{18393375}\right) p_\infty^{13}\nonumber\\
&&
+\left(\frac{178687772947
   \ln(p_\infty/2)}{2536934400}-\frac{13258067 \pi^2}{1182720}-\frac{23690615024526544067}{108790951965696000}\right) p_\infty^{14}
+\left(\frac{43153771264}{70945875}-\frac{153233329 \pi^2}{11714560}\right) p_\infty^{15}\nonumber\\
&&\left.
+\left(-\frac{8273777127559
   \ln(p_\infty/2)}{102604902400}+\frac{220334427 \pi^2}{20500480}+\frac{19805045466762324597221}{95542631592984576000}\right) p_\infty^{16}
+O(p_\infty^{17})\right]
\,.\nonumber\\
\eea
It agrees with the small-velocity expansion of the exact result of Ref. \cite{Driesse:2024feo}.

The 4PM-1SF radiated angular momentum turns out to be 
\bea
J_4^{\rm 1SF} &=& \frac{176}{5} p_\infty+\frac{8144}{105} p_\infty^3+\frac{448}{5} p_\infty^4-\frac{93664}{1575} p_\infty^5
+\frac{1184}{21} p_\infty^6-\frac{4955072}{121275} p_\infty^7-\frac{13736}{315} p_\infty^8\nonumber\\
&-&\frac{29857664}{1576575}p_\infty^9+\frac{724868}{17325} p_\infty^{10}+\frac{28280064}{1926925} p_\infty^{11}
-\frac{15578279}{450450}p_\infty^{12}-\frac{551268352}{32757725}p_\infty^{13}+\frac{20316617}{700700} p_\infty^{14}\nonumber\\
&+& \frac{4799560603648}{218461268025}p_\infty^{15}+O(p_\infty^{16})
\,.
\eea
The even powers of $p_\infty$ agree with the PN expansion of the corresponding exact expression given in Ref. \cite{Heissenberg:2025ocy}.
The remaining terms instead are new with this work.

The 5PM contribution to the radiation-reacted scattering angle \eqref{chirad} is then (see Eq. \eqref{chiraddef} for the definition of the coefficients $\chi^{\rm rad}_n$)
\beq
\chi^{\rm rad}_5=\nu\chi^{\rm rad,\,1SF}_5+O(\nu^2)\,,
\eeq
with
\bea
\chi^{\rm rad,\,1SF}_5&=&
\frac{1504}{45}
+\left(\frac{43184}{525}+\frac{42 \pi ^2}{5}\right) p_\infty^2
+\frac{3584}{45} p_\infty^3
+\left(\frac{1741664}{11025}+\frac{267 \pi^2}{14}\right) p_\infty^4
+\frac{2496}{35} p_\infty^5
+\left(\frac{541568}{3465}+\frac{51847 \pi^2}{6720}\right) p_\infty^6\nonumber\\
&&
+\frac{37664}{1575} p_\infty^7
+\left(\frac{94791296}{3468465}-\frac{1786689 \pi^2}{197120}\right) p_\infty^8
-\frac{1845832}{17325} p_\infty^9
+\left(\frac{22838308096}{676350675}+\frac{34854341 \pi ^2}{20500480}\right) p_\infty^{10}\nonumber\\
&&
+\frac{87905096}{1576575} p_\infty^{11}
+\left(-\frac{2121720175616}{34493884425}-\frac{93027497 \pi ^2}{123002880}\right) p_\infty^{12}
-\frac{513823}{14014} p_\infty^{13}\nonumber\\
&&
+\left(\frac{21467916673024}{247589437095}+\frac{39897833 \pi^2}{99573760}\right) p_\infty^{14}
+O(p_\infty^{15})
\,.
\eea

\section{Conclusions}

The radiated energy and angular momentum emitted by a point mass scattered by a Schwarzschild black hole have been analytically computed in the framework of first-order self-force theory in the form of combined PM-PN expansions, up to 5PM for the energy and 4PM for the angular momentum, and 7PN for both.
Two recent works have computed the same quantities through different methods, Refs. \cite{Driesse:2024feo} and \cite{Heissenberg:2025ocy}.
The 5PM energy agrees with the PN expansion of the exact expression of Ref. \cite{Driesse:2024feo}.
The 4PM angular momentum has been determined instead only partially in Ref. \cite{Heissenberg:2025ocy}, and completed here up to the 7PN order.
The knowledge of the full 4PM radiated angular momentum has also allowed for determining the 5PM-1SF contribution to the radiation-reacted scattering angle, which should serve as a useful check for similar calculations by other methods.
Going beyond the 5PM level for the energy (and 4PM for the angular momentum) would require a further significant effort, either in terms of computational time and calculation of nolocal integrals \eqref{nlint}, which will involve multiple polylogarithms.
This is left for future work.

\section*{Acknowledgments}
I would like to thank Thibault Damour for valuable comments on a preliminary version of this manuscript.
I also thank Donato Bini and Davide Usseglio for useful discussions. 
I'm grateful to Mario Vasile for his kind assistance in running from remote part of the codes related with this work, so speeding up the accomplishment of the final result.

\end{document}